\documentclass[11pt,cls,onecolumn]{IEEEtran}
\usepackage{graphicx, amsmath, amssymb}
\begin{document}

\title{Capacity Scaling of Single-source Wireless Networks: Effect of Multiple Antennas}
\author{Sang-Woon Jeon,~\IEEEmembership{Student Member,~IEEE,}
        and~Sae-Young~Chung,~\IEEEmembership{Senior Member,~IEEE\\}
\authorblockA{School of EECS, KAIST, Daejeon, Korea\\}
Email: swjeon@kaist.ac.kr, sychung@ee.kaist.ac.kr}
\maketitle


\newtheorem{definition}{Definition}
\newtheorem{theorem}{Theorem}
\newtheorem{lemma}{Lemma}
\newtheorem{example}{Example}
\newtheorem{corollary}{Corollary}
\newtheorem{proposition}{Proposition}
\newtheorem{conjecture}{Conjecture}
\newtheorem{remark}{Remark}

\def \diag{\operatornamewithlimits{diag}}
\def \min{\operatornamewithlimits{min}}
\def \max{\operatornamewithlimits{max}}
\def \log{\operatorname{log}}
\def \max{\operatorname{max}}
\def \rank{\operatorname{rank}}
\def \out{\operatorname{out}}
\def \exp{\operatorname{exp}}
\def \arg{\operatorname{arg}}
\def \E{\operatorname{E}}
\def \tr{\operatorname{tr}}
\def \SNR{\operatorname{SNR}}
\def \SINR{\operatorname{SINR}}
\def \dB{\operatorname{dB}}
\def \ln{\operatorname{ln}}
\def \th{\operatorname{th}}

\begin{abstract}
We consider a wireless network in which a single source node located at the center of a unit area having $m$ antennas transmits messages to $n$ randomly located destination nodes in the same area having a single antenna each.
To achieve the sum-rate proportional to $m$ by transmit beamforming, channel state information (CSI) is essentially required at the transmitter (CSIT), which is hard to obtain in practice because of the time-varying nature of the channels and feedback overhead.
We show that, even without CSIT, the achievable sum-rate scales as $\Theta(m\log m)$ if a cooperation between receivers is allowed.
By deriving the cut-set upper bound, we also show that $\Theta(m\log m)$ scaling is optimal.
Specifically, for $n=\omega(m^2)$, the simple TDMA-based quantize-and-forward is enough to achieve the capacity scaling.
For $n=\omega(m)$ and $n=\operatorname{O}(m^2)$, on the other hand, we apply the hierarchical cooperation to achieve the capacity scaling.
\end{abstract}


\section{Introduction} \label{sec:intro}
In a pioneering work of \cite{GuptaKumar:00}, Gupta and Kumar have studied the sum-rate scaling of wireless ad hoc networks as the number $n$ of randomly located source-destination (S-D) pairs increases in a fixed area.
They showed that the sum-rate scales as $\Theta\left(\sqrt{n/\log n}\right)$ using a nearest multihop transmission.
The hierarchical cooperation scheme was recently proposed in \cite{OzgurLevequeTse:06} improving the sum-rate scaling drastically.
Specifically, the sum-rate scales almost linearly in $n$ that is well matched with the information-theoretic upper bound.
Then a natural question is how the sum-rate scales if the number of sources and the number of destinations are not balanced.
As extreme cases, we can consider a single source sending messages to the rest of the nodes or a single destination collecting messages from the rest of the nodes in the network.
The work in \cite{HeshamElGamal:05} has considered the single-destination network and proved that the capacity scales as $\Theta\left(\log n\right)$, which is also true for the single-source network.
Notice that this result indicates that adding more nodes in the network only provides a marginal gain, that is, the per-node rate tends to zero.

One of the promising approaches to resolve the unbalanced network topology is to adopt multiple antennas at the source (or at the destination).
The works in \cite{FoschiniGans:98, Telatar:99} show that having multiple antennas increases the capacity of a point-to-point link proportionally to the minimum number of transmit and receive antennas when the channel state information (CSI) is available at the receiver.
This result can be directly extended to the single-destination network in which the destination has $m$ antennas and $n$ sources have a single antenna each.
Thus, the sum-rate increases proportionally to $m$ if $n=\Omega(m)$.
For the single-source network, achieving linear scaling becomes much more challenging because the source should acquire CSI in order to apply the transmit beamforming techniques such as the dirty paper coding with optimal beamforming \cite{Caire:03, Weingarten:06} or zero-forcing beamforming \cite{Yoo:06}. 
From a practical point of view, it is hard to obtain CSI at the transmitter (CSIT) due to the time-varying channels and feedback overhead, especially for a large wireless network with many antennas at the source.

In this paper, we study the capacity scaling of the single-source network having multiple antennas at the source when CSI is available only at the receiver side.
We mainly focus on the feasibility of the linear increase of the sum-rate proportional to the number of source antennas by using receiver cooperation only.
Thus, we address the questions regarding the minimum number of required nodes in the network to achieve this linear scaling and the type of receiver cooperation that can achieve the optimal scaling law.
To answer these questions, we first derive an information-theoretic upper bound and propose a cooperative multiple-input multiple-output (MIMO) scheme achieving the same scaling as the upper bound.

\section{System Model} \label{sec:sys_model}
In this section, we define the underlying network and channel models and explain the performance metric used in the paper.
The matrix and vector operations used in the paper are summarized in Table \ref{Table:simbols}.

\subsection{Single-source Wireless Networks}
We consider a single-source wireless network.
There exists a single source having $m$ antennas at the center of the network and $n$ destinations each having a single antenna are uniformly distributed over a unit square area.
The relation between $m$ and $n$ is given by
\begin{equation}
n=m^{\beta},
\end{equation}
where $\beta>0$.
Let $\mathbf{r}_0$ denote the position of the source and $\mathbf{r}_{i}$ denote the position of the $i$-th destination, where $i\in\{1,\cdots,n\}$.

We consider a far-field channel model.
The $1\times m$ channel vector from the source to the $i$-th destination at time $t$ is given by
\begin{equation}
\mathbf{h}_{0i}(t)={\|\mathbf{r}_0-\mathbf{r}_i\|}^{-\alpha/2}\big[e^{j2\pi \theta_{0i,1}(t)},\cdots, e^{j2\pi \theta_{0i,m}(t)}\big]\mbox{ for }i\in\{1,\cdots,n\},
\label{EQ:h_0i}
\end{equation}
where $\alpha>2$ is the path-loss exponent and $\theta_{0i,j}(t)$ is the phase at time $t$ from the $j$-th transmit antenna of the source to the $i$-th destination.
Now consider the channel from the $i$-th to the $j$-th destinations at time $t$ that is given by
\begin{equation}
h_{ij}(t)={\|\mathbf{r}_i-\mathbf{r}_j\|}^{-\alpha/2}e^{j2\pi \theta_{ij}(t)}\mbox{ for }i\neq j \mbox{ and } i, j\in\{1,\cdots,n\},
\end{equation}
where $\theta_{ij}(t)$ is the phase at time $t$ from the $i$-th to the $j$-th destinations.
We assume fast fading in which $\theta_{0i,j}(t)$ and $\theta_{kl}(t)$ are uniformly distributed within $[0,2\pi)$ independent for different $i$, $j$, $k$, $l$, and $t$.
We further assume that CSI is available only at the receivers.

The received signal of the $j$-th destination at time $t$ is given by
\begin{equation}
y_j(t)=\mathbf{h}_{0j}(t)\mathbf{x}_0(t)+\sum_{i=1,i\neq j}^n h_{ij}(t)x_i(t)+z_j(t)\mbox{ for }j\in\{1,\cdots,n\},
\label{EQ:input_output}
\end{equation}
where $\mathbf{x}_0(t)$ is the $m\times 1$ transmit signal vector of the source, $x_i(t)$ is the transmit signal of the $i$-th destination, and $z_i\sim\mathcal{N}_{\mathbb{C}}(0,1)$ is the noise of the $i$-th destination that is independent for different $i$ and $t$.
Note that not all destinations transmit simultaneously at a given time and some $x_i(t)$'s in (\ref{EQ:input_output}) may be zero.
The source and each destination have power constraints $P_0$ and $P_1$, respectively. Thus $\mathbb{E}\left(\|\mathbf{x}_0(t)\|^2\right)\leq P_0$ and $\mathbb{E}\left(|x_i(t)|^2\right)\leq P_1$ for $i\in\{1,\cdots,n\}$.
For notational convenience, we will omit time index $t$ in the rest of the paper.

\subsection{Performance Measure}
Throughout the paper, we will analyze the capacity scaling of the single-source wireless network, which could be a function of $m$ and $n$.
We define the individual rate $R_{\operatorname{ind}}$ such that for suffuciently large $m$ the source can transmit with at least $R_{\operatorname{ind}}$ bps/Hz to each of $n$ destinations with high probability (whp).
Then the achievable sum-rate is simply given by $R_{\operatorname{sum}}=nR_{\operatorname{ind}}$.
For notational simplicity, `whp' used in the paper means that an event occurs with high probability as $m\to \infty$\footnote{Since $n=m^{\beta}$, $n$ also tends to infinity as $m\to \infty$.}.

%

\section{Cut-Set Upper Bound} \label{sec:upper_bound}
In this section, we obtain an upper bound on the sum-rate, which will be compared to the achievable sum-rate derived in the next section.
Let $\mathbf{H}_0$ denote the $n\times m$ compound MIMO channel from the source to the $n$ destinations that is given by
\begin{equation}
\mathbf{H}_0=[\mathbf{h}_{01}^T,\cdots,\mathbf{h}_{0n}^T]^T.
\end{equation}
From the cut-set bound, we know that the sum-rate is upper bounded by the rate across the cut dividing the source and the $n$ destinations.
By assuming full cooperation between the destinations, the sum-rate is upper bounded by MIMO capacity.
Thus, we obtain
\begin{equation}
R_{\operatorname{sum}}\leq\underset{\operatorname{Tr}(\mathbf{\Sigma}_0)\leq 1}{\max}\mathbb{E}\left(\log \det \left(\mathbf{I}_n+P_0\mathbf{H}_0\mathbf{\Sigma}_0\mathbf{H}_0^\dagger\right)\right),
\label{EQ:ergodic_cap}
\end{equation}
where $\mathbf{\Sigma}_0$ denotes the $m\times m$ normalized input covariance matrix given by $\frac{1}{P_0}\mathbb{E}(\mathbf{x}_0\mathbf{x}_0^{\dagger})$.

\begin{theorem} \label{THM:rate_upper}
Suppose the single-source wireless network. For sufficiently large $m$, $R_{\operatorname{sum}}$ scales as $\operatorname{O}(m^{\min\{1,\beta\}}\log m)$ whp.
\end{theorem}

\begin{proof}
For $\beta>1$, from (\ref{EQ:ergodic_cap}), we obtain
\begin{eqnarray}
R_{\operatorname{sum}}\!\!\!\!\!\!\!\!\!\!&&\overset{(a)}{\leq}\underset{\operatorname{Tr}(\mathbf{\Sigma}_0)\leq 1}{\max}\mathbb{E}\left(\log \det \left(\mathbf{I}_m+P_0\mathbf{H}_0^\dagger\mathbf{H}_0\mathbf{\Sigma}_0\right)\right)\nonumber\\
&&\overset{(b)}{\leq}\underset{\operatorname{Tr}(\mathbf{\Sigma}_0)\leq 1}{\max}\log \det \left(\mathbf{I}_m+P_0\mathbb{E}(\mathbf{H}_0^\dagger\mathbf{H}_0)\mathbf{\Sigma}_0\right)\nonumber\\
&&\overset{(c)}{=}\underset{\operatorname{Tr}(\mathbf{\Sigma}_0)\leq 1}{\max}\log \det \left(\mathbf{I}_m+P_0\sum_{i=1}^{n}\|\mathbf{r}_0-\mathbf{r}_i\|^{-\alpha}\mathbf{\Sigma}_0\right)\nonumber\\
&&\overset{(d)}{=}m\log\left(1+\frac{P_0}{m}\sum_{i=1}^{n}\|\mathbf{r}_0-\mathbf{r}_i\|^{-\alpha}\right).
\end{eqnarray}
Notice that $(a)$ holds since $\det\left(\mathbf{I}_n+\mathbf{A}\mathbf{A}^\dagger\right)=\det\left(\mathbf{I}_m+\mathbf{A}^\dagger\mathbf{A}\right)$, $(b)$ holds since $\log\det(\cdot)$ is concave \cite{Boyd:04}, $(c)$ holds since $\mathbb{E}(\mathbf{H}_0^\dagger\mathbf{H}_0)=\sum_{i=1}^{n}\|\mathbf{r}_0-\mathbf{r}_i\|^{-\alpha}\mathbf{I}_m$, and $(d)$ holds since the rate is maximized by $\mathbf{\Sigma}_0=\frac{1}{m}\mathbf{I}_m$ \cite{Telatar:99}.
From the fact that the minimum distance between the source and any other destination is larger than $n^{-(1+\epsilon_d)}$ whp, where $\epsilon_d>0$ is an arbitrarily small constant \cite{OzgurLevequeTse:06}, we finally obtain $R_{\operatorname{sum}}\leq m\log\left(1+P_0 m^{\beta+\alpha\beta(1+\epsilon_d)-1}\right)$ whp.
Thus the upper bound scales as $\Theta(m\log m)$ whp.

For $0<\beta\leq 1$, we obtain
\begin{eqnarray}
R_{\operatorname{sum}}\!\!\!\!\!\!\!\!\!\!&&\leq\sum_{i=1}^n \underset{\operatorname{Tr}(\mathbf{\Sigma}_0)\leq 1}{\max}\mathbb{E}\left(\log\left(1+P_0\mathbf{h}_{0i}\mathbf{\Sigma}_0\mathbf{h}_{0i}^{\dagger}\right)\right)\nonumber\\
&&\leq\sum_{i=1}^n\mathbb{E}\left(\log\left(1+P_0\mathbf{h}_{0i}\mathbf{h}^{\dagger}_{0i}\right)\right)\nonumber\\
&&=n\log\left(1+P_0m\|\mathbf{r}_0-\mathbf{r}_i\|^{-\alpha}\right),
\end{eqnarray}
where we use generalized Hadamard's inequality for the first inequality and set $\mathbf{\Sigma}_0=\mathbf{I}_m$ for the second inequality.
Thus $R_{\operatorname{sum}}\leq m^{\beta}\log\left(1+P_0m^{1+\alpha\beta(1+\epsilon_d)}\right)$ whp that scales as $\Theta(m^{\beta}\log m)$.
Therefore, Theorem \ref{THM:rate_upper} holds.
\end{proof}

\section{Quantize-And-Forward-Based Cooperative MIMO} \label{sec:proposed}
In this section, we propose a cooperative MIMO scheme and analyze its achievable rate.
We define a small region around the source, and serve the destinations in that region using a small fraction of time.
Notice that it is possible to set the area of the region and time fraction such that it does not affect the overall rate scaling while making the distance between the source and the destinations outside the region as $\Theta(1)$.
Thus the proposed scheme is about the transmission to the destinations outside the region.
For simplicity, we assume all destinations are located outside the region in the rest of the paper.

\begin{figure}[t!]
  \begin{center}
  \scalebox{0.7}{\includegraphics{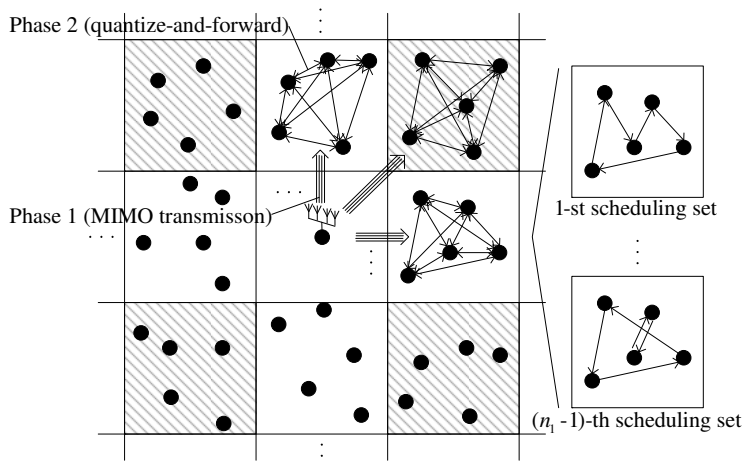}}
  \caption{The overall procedure of the proposed scheme, where the group in the shaded cells denote the active groups determined by $4$-TDMA.}
  \label{FIG:proposed_scheme}
  \end{center}
\end{figure}

As mentioned before, due to the lack of CSI, the source cannot perform a coherent beamforming.
Instead, we apply the cooperative MIMO to induce the cooperation among the destinations.
We first divide the entire destinations into several groups and perform cooperative MIMO between the source and each group.
Let $n_1$ denote the number of groups and $n_2$ denote the number of destinations in each group, where $n=n_1n_2$.
We denote the $i$-th destination in the $k$-th group as destination $(k,i)$, where $i\in\{1,\cdots,n_2\}$ and $k\in\{1,\cdots,n_1\}$.
Denote the positions of $n_2$ destinations in the $k$-th group as $\left\{\mathbf{r}_1^{k},\cdots,\mathbf{r}_{n_2}^{k}\right\}$.
Without loss of generality, we assume $\|\mathbf{r}_0-\mathbf{r}^k_i\|\leq \|\mathbf{r}_0-\mathbf{r}^k_j\|$ for $i<j$.
Now consider the message transmission from the source to destination $(k,i)$.
The proposed scheme consists of two phases.
In Phase $1$, the source transmits a message to destination $(k,i)$ by using the other nodes in the $k$-th group as relays.
To relay the received signals, the other nodes quantize their received signals and transmit them to destination $(k,i)$, which is Phase $2$.
Destination $(k,i)$ finally decodes the message based on these quantized signals.
We apply the TDMA-based cooperation or the hierarchical cooperation in \cite{OzgurLevequeTse:06} for relaying the quantized received signals within each group.
Fig. \ref{FIG:proposed_scheme} illustrates the overall procedure of the proposed cooperative MIMO scheme, where the destinations in the same cell will form a group.
We will explain the details of Phases $1$ and $2$ in the next two subsections.

\subsection{Phase 1: MIMO Transmission}
Using the $\Delta$ fraction of time, Phase $1$ performs the following procedure.
\begin{itemize}
\item $n_1$-TDMA is used among $n_1$ groups.
\item $n_2$-TDMA is used among $n_2$ destinations in the same group.
\item The source transmits the message of destination $(k,i)$ via Gaussian signaling with covariance matrix $\frac{P_0\|\mathbf{r}_0-\mathbf{r}^k_{n_2}\|^{\alpha}}{m}\mathbf{I}_m$ to the nodes in the $k$-th group.
\end{itemize}

Since only the source transmits in Phase $1$, from (\ref{EQ:input_output}), the received signal of destination $(k,j)$ in Phase $1$ is given by
\begin{equation}
y^k_{P1,j}=\mathbf{h}^k_{0j}\mathbf{x}^k_0+z^k_j=\sqrt{\frac{P_0\|\mathbf{r}_0-\mathbf{r}^k_{n_2}\|^{\alpha}}{m}}\mathbf{h}^k_{0j}\mathbf{s}^k_0+z^k_j,
\label{EQ:input_output_p1}
\end{equation}
where the superscript denotes the group index and $\mathbf{s}^k_0$ follows a complex Gaussian distribution with $\mathcal{N}_{\mathbb{C}}(0,\mathbf{I}_m)$.
From (\ref{EQ:h_0i}), $\mathbf{h}^k_{0i}$ is given by $\|\mathbf{r}_0-\mathbf{r}^k_i\|^{-\alpha/2}\big[e^{j2\pi \theta^k_{0i,1}},\cdots, e^{j2\pi \theta^k_{0i,m}}\big]$, where $\theta^k_{0i,j}$ is the phase from the $j$-th antenna of the source to destination $(k,i)$.
Let $\mathbf{H}^k_0=\left[\mathbf{h}^{kT}_{01},\cdots,\mathbf{h}^{kT}_{0n_2}\right]^T$, $\mathbf{y}^k_{P1}=\left[y_{P1,1}^k,\cdots,y_{P1,n_2}^k\right]^T$, and $\mathbf{z}^k_0=\left[z^k_{01},\cdots, z^k_{0n_2}\right]^T$.
Then
\begin{equation}
\mathbf{y}_{P1}^k=\mathbf{H}^k_0\mathbf{x}^k_0+\mathbf{z}^k_0=\sqrt{\frac{P_0}{m}}\mathbf{\Gamma}^{k}_0\mathbf{\Theta}^k_0\mathbf{s}^k_0+\mathbf{z}_0^k,
\label{EQ:y_kP1}
\end{equation}
where $\mathbf{\Gamma}^k_0=\diag\left(\frac{\|\mathbf{r}-\mathbf{r}^k_{n_2}\|^{\frac{\alpha}{2}}}{\|\mathbf{r}-\mathbf{r}^k_1\|^{\frac{\alpha}{2}}},\cdots,\frac{\|\mathbf{r}-\mathbf{r}^k_{n_2}\|^{\frac{\alpha}{2}}}{\|\mathbf{r}-\mathbf{r}^k_{n_2}\|^{\frac{\alpha}{2}}}\right)$ and $\left[\mathbf{\Theta}^k_0\right]_{ij}=e^{j2\pi\theta^{k}_{0i,j}}$.
Although  the Gaussian signaling may not be optimal because the elements of $\mathbf{H}^k_0$ are not $i.i.d.$ \cite{Telatar:99}, we will show that it achieves the optimal capacity scaling in the next section.

\subsection{Phase 2: Quantize-and-Forward}
Because each destination should collect the quantized received signals from the other $n_2-1$ nodes in the same group, $n_2(n_2-1)$ transmission pairs should be served in each group during Phase $2$.
Notice that by choosing $n_2$ transmission pairs such that each of $n_2$ nodes becomes a transmitter and a receiver of two different pairs, we can construct $n_2-1$ scheduling sets as shown in Fig. \ref{FIG:proposed_scheme}.
Now consider the transmission of $n_2$ pairs in each scheduling set.
For the communcation, we can use transmission schemes proposed for the ad hoc network model, for example the hierarchical cooperation in \cite{OzgurLevequeTse:06}, as well as the simple TDMA to serve the pairs in each scheduling set.
Using the $1-\Delta$ fraction of time, Phase $2$ performs the following procedure.
\begin{itemize}
\item $4$-TDMA is used among adjacent groups, that is, one out of $4$ groups are activated simultaneously.
\item $(n_2-1)$-TDMA is used among $(n_2-1)$ scheduling sets in the same group.
\item One of the following two is performed.

1. TDMA-based cooperation: $n_2$-TDMA is used among $n_2$ transmission pairs in the same scheduling set.

2. Hierarchical cooperation: Hierarchical cooperation is used among $n_2$ transmission pairs in the same scheduling set.
\item For each transmission pair, the transmitter quantizes its received signal and transmits it to the receiver with power $P_1$.
\end{itemize}
\vspace{0.1in}


Let $h^k_{ij}$ denote the channel from destination $(k,i)$ to destination $(k,j)$ and $h^{lk}_{ij}$ denote the channel from destination $(l,i)$ to destination $(k,j)$.
Suppose that $\mathcal{A}(t)$ is the set of active groups at time $t$, which is determined by $4$-TDMA.
Then the received signal of destination $(k,j)$ when destination $(k,i)$ transmits is given by
\begin{eqnarray} \label{EQ:received_signal_p2}
y_{P2,j}^k=h^k_{ij}x^k_i+\sum_{l\in\mathcal{A},l\neq k}h^{lk}_{ij}x^l_i+z^k_j,
\end{eqnarray}
where we assume that the $i$-th destination also transmits in the other active groups.

\section{Performance Analysis}
In this section, we derive an achievable rate of the proposed scheme and analyze its scaling law.

\subsection{Achievable Rate}
\begin{figure}[t!]
  \begin{center}
  \scalebox{0.7}{\includegraphics{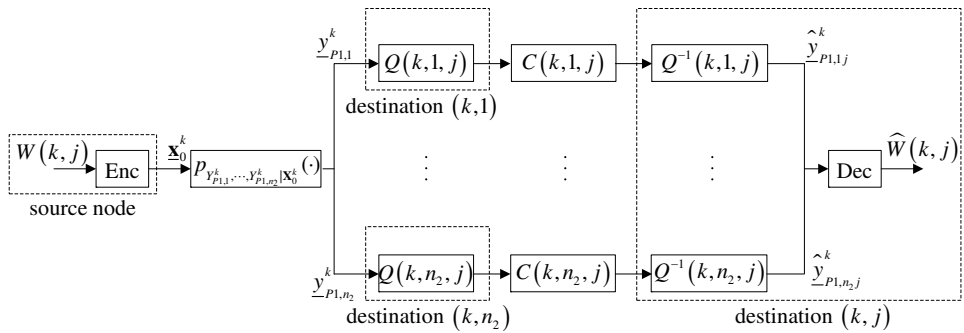}}
  \caption{Cooperative MIMO from the source to destination $(k,j)$.}
  \label{FIG:quantized_MIMO}
  \end{center}
\end{figure}

Fig. \ref{FIG:quantized_MIMO} illustrates the message transmission from the source to destination $(k,j)$.
Let $T$ denote the block length and $\underline{y}^k_{P1,i}$ denote $[y^k_{P1,i}(1),\cdots,y^k_{P1,i}(T)]$.
In Phase 1, the source sends a message $W(k,j)$ to the destinations in the $k$-th group through the channel $p_{Y^k_{P1,1},\cdots,Y^k_{P1,n_2}|\mathbf{X}^k_0}(\cdot)$.
In Phase 2, destination $(k,i)$ quantizes its received signal $\underline{y}^k_{P1,i}$ and transmits the quantized signal $\underline{\hat{y}}^k_{P1,ij}$ to destination $(k,j)$ through the link having a finite capacity of $C(k,i,j)$ for all $i\in\{1,\cdots,n_2\}$.
Destination $(k,j)$ finally decodes $W(k,j)$ based on the quantized outputs $\underline{\hat{y}}^k_{P1,1j}, \cdots, \underline{\hat{y}}^k_{P1,n_2j}$.
Therefore, the source and destination $(k,j)$ have $(2^{TR(k,j)};T)$ channel encoder and decoder, respectively, where $R(k,j)$ denotes the data rate.
Destination $(k,i)$ and destination $(k,j)$ have $(2^{TR_Q(k,i,j)};T)$ quantizer and dequantizer, respectively, where $R_Q(k,i,j)$ denotes the quantization rate.
The achievable rates can be derived by modifying the result in \cite{OzgurLevequeTse:06}.

\begin{theorem}[{\"O}zg{\"u}r, L{\'e}v{\^e}que, and Tse] \label{THM:ozgur}
Given a probability distribution $p_{\mathbf{X}^k_0}(\cdot)$ and  $n_2$ conditional probability distributions $p_{\hat{Y}^k_{P1,ij}|Y^k_{P1,i}}(\cdot)$ for $i\in\{1,\cdots,n_2\}$, the rates satisfying
\begin{eqnarray}
R_Q(k,i,j)\!\!\!\!\!\!\!\!\!&&\leq\frac{1-\Delta}{4n_2}C(k,i,j),{~} i\in\{1,\cdots,n_2\}\label{EQ:R_21}\\
R_Q(k,i,j)\!\!\!\!\!\!\!\!\!&&\geq\frac{\Delta}{n}I(Y^k_{P1,i};\hat{Y}^k_{P1,ij}),{~} i\in\{1,\cdots,n_2\}\label{EQ:R_22}\\
R(k,j)\!\!\!\!\!\!\!\!\!&&\leq\frac{\Delta}{n} I(\mathbf{X}_0^k;\hat{Y}^k_{P1,1j},\cdots,\hat{Y}^k_{P1,n_2j})\label{EQ:R}
\end{eqnarray}
are achievable.
\end{theorem}

\begin{proof}
The constraints (\ref{EQ:R_22}) and (\ref{EQ:R}) are directly obtained from Theorem II.1 in \cite{OzgurLevequeTse:06} by multiplying $\frac{\Delta}{n}$ because each destination is served using $\frac{\Delta}{n}$ time fraction in Phase $1$.
The first constraint comes from the fact that each quantizer should deliver its quantization index through a finite capacity link using $\frac{1-\Delta}{4n_2}$ time fraction in Phase $2$, where $\frac{1}{4}$ reflects the effect of $4$-TDMA.
\end{proof}

Now consider the Gaussian channel in (\ref{EQ:input_output_p1}) and (\ref{EQ:y_kP1}).
Let $N^k_{ij}$ be the variance of the quantization noise given by $N^k_{ij}=\mathbb{E}(|Y^k_{P1,i}-\hat{Y}^k_{P1,ij}|^2)$.
If we set the conditional probability distributions as $p_{\hat{Y}^k_{P1,ij}|Y^k_{P1,i}}(\cdot)\sim \mathcal{N}_{\mathbb{C}}(y^k_{P1,i},N^k_{ij})$ for all $i\in\{1,\cdots,n_2\}$, then
\begin{equation} \label{EQ:condition_r2}
I(Y^k_{P1,i};\hat{Y}^k_{P1,ij})\leq \log\left(1+\frac{\mathbb{E}(|Y^k_{P1,i}|^2)}{N^k_{ij}}\right),
\end{equation}
where we use the fact that the Gaussian distribution maximizes entropy for a given received power \cite{Cover:91}.
From (\ref{EQ:R_21}) and (\ref{EQ:R_22}), if the following condition is satisfied
\begin{equation}
\frac{1-\Delta}{4n_2}C(k,i,j)\geq \frac{\Delta}{n}\log \left(1+\frac{\mathbb{E}(|Y^k_{P1,i}|^2)}{N_{ij}^k}\right) \mbox{ for all } i\in\{1,\cdots,n_2\},
\label{EQ:condition_R2}
\end{equation}
then we can find $R(k,i,j)$ satisfying (\ref{EQ:R_21}) and (\ref{EQ:R_22}) simultaneously.
Thus we set $N^k_{ij}$ as the minimum value that satisfies (\ref{EQ:condition_R2}) with equality such that
\begin{equation}
N^k_{ij}=\frac{\mathbb{E}(|Y^k_{P1,i}|^2)}{2^{\frac{1-\Delta}{\Delta}\frac{n}{4n_2}C(k,i,j)}-1}.
\label{EQ:Quan_noise}
\end{equation}
Let $\hat{\mathbf{y}}^k_{P1,j}=[\hat{y}_{P1,1j},\cdots,\hat{y}_{P1,n_2j}]^T$. Then $\hat{\mathbf{y}}^k_{P1,j}=\mathbf{H}^k_0\mathbf{x}^k_0+\mathbf{z}^k_0+\hat{\mathbf{z}}^k_{0j}$, where $\mathbb{E}(\hat{\mathbf{z}}^k_{0j}\hat{\mathbf{z}}^{k\dagger}_{0j})=\diag(N^k_{1j},\cdots,N^k_{n_2j})$.
Therefore, from (\ref{EQ:R}), we conclude
\begin{equation}
R(k,j)=\frac{\Delta}{n}\mathbb{E}\left(\log\det\left(\mathbf{I}_{n_2}+\frac{P_0}{m}\mathbf{\Gamma}^k_0\mathbf{\Theta}^k_0\mathbf{\Theta}^{k\dagger}_0\mathbf{\Gamma}^k_0(\mathbf{Q}^k_j)^{-1}\right)\right)
\label{EQ:R_kj}
\end{equation}
is achievable, where $\mathbf{Q}^k_j=\diag(1+N^k_{1j},\cdots,1+N^k_{n_2j})$.

There exists a trade-off between the size of MIMO and the variance of the quantization noise $N^k_{ij}$.
If we set $n_2$ as a small value from (\ref{EQ:Quan_noise}), we can make $N^k_{ij}$ small.
The reason is that as $n_2$ decreases, that is $n_1$ increases, we have more spatially reusable groups in Phase $2$ so that the aggregate rate of Phase $2$ increases and, as a result, we can decrease $N^k_{ij}$.
Although small $n_2$ has an advantage in terms of quantization noises, which increases $R(k,j)$, it also decreases the size of $\mathbf{H}^k_0$, which decreases $R(k,j)$.
In the next subsection, we will show that the optimal rate scaling is achievable by choosing $n_1$ and $n_2$ properly.

\subsection{Asymptotic Analysis}
In this subsection, we derive $R_{\operatorname{sum}}$ of the proposed scheme in the limit of large $m$.
To specify $n_1$ and $n_2$, we divide the network into small square cells of area $n^{-q}$, where $q\in(0,1)$.
Then $n_1$ is given by $n^q$ and $n_2$ is approximately given by $n^{1-q}$ whp, which will be proved in Lemma \ref{THM:n_2_value}.
Before deriving the main results, we consider the following lemmas, which will be used to prove the main results.

\begin{lemma} \label{THM:n_2_value}
For sufficiently large $m$, the number of destinations in any cell is in $[(1-\delta)n^{1-q},(1+\delta)n^{1-q}]$ whp, where $\delta>0$ is an arbitrarily small constant.
\end{lemma}
\begin{proof}
We refer Lemma $4.1$ in \cite{OzgurLevequeTse:06} for the proof.
\end{proof}

\begin{lemma} \label{THM:max_pwr}
For sufficiently large $m$, $\mathbb{E}(|Y^k_{P1,i}|^2)$ is upper bounded by $P_0(1+\epsilon_p)+1$ whp for all $i$ and $k$, where $\epsilon_p>0$ is an arbitrarily small constant.
\end{lemma}
\begin{proof}
From (\ref{EQ:input_output_p1}), we obtain
\begin{eqnarray}
\mathbb{E}(|Y^k_{P1,i}|^2)\!\!\!\!\!\!\!\!\!&&=\mathbf{h}^k_{0i}\mathbb{E}(\mathbf{x}_0^k\mathbf{x}_0^{k\dagger})\mathbf{h}^{k\dagger}_{0i}+1\nonumber\\
&&=P_0\frac{\|\mathbf{r}_0-\mathbf{r}^k_{n_2}\|^{\alpha}}{\|\mathbf{r}_0-\mathbf{r}^k_{i}\|^{\alpha}}+1\\
&&=P_0\left(1+\frac{\|\mathbf{r}_0-\mathbf{r}^k_{n_2}\|-\|\mathbf{r}_0-\mathbf{r}^k_i\|}{\|\mathbf{r}_0-\mathbf{r}^k_i\|}\right)^{\alpha}+1\nonumber\\
&&\leq P_0\left(1+\frac{\|\mathbf{r}_{n_2}-\mathbf{r}^k_i\|}{\|\mathbf{r}_0-\mathbf{r}^k_i\|}\right)^{\alpha}+1\nonumber\\
&&\leq P_0\left(1+\frac{\sqrt{2}n^{-\frac{q}{2}}}{\|\mathbf{r}_0-\mathbf{r}^k_i\|}\right)^{\alpha}+1,
\end{eqnarray}
where the first inequality holds from the triangular inequality and the second inequality holds since both destinations $(k,i)$ and $(k,n_2)$ are placed in the same square cell of area $n^{-q}$.
Because $\|\mathbf{r}_0-\mathbf{r}^k_i\|=\Theta(1)$ whereas $n^{-\frac{q}{2}}\to 0$ as $n\to\infty$, $\mathbb{E}(|Y^k_{P1,i}|^2)$ is upper bounded by $P_0(1+\epsilon_p)+1$ whp, where $\epsilon_p>0$ is an arbitrarily small constant.
Thus Lemma \ref{THM:max_pwr} holds.
\end{proof}

\begin{figure}[t!]
  \begin{center}
  \scalebox{0.8}{\includegraphics{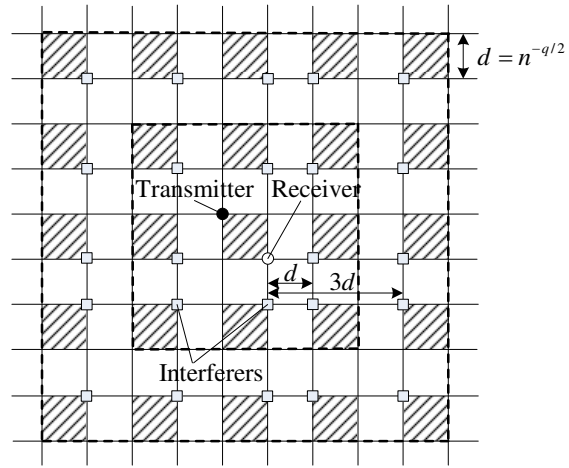}}
  \caption{Worst interference for Phase $2$.}
  \label{FIG:worst_sinr}
  \end{center}
\end{figure}

\begin{lemma} \label{THM:rate_p2_tdma}
Suppose that the TDMA-based cooperation is used within a group.
For sufficiently large $m$, $C(k,i,j)$ is lower bounded by $C_1n^{-1}_2$, where $C_1>0$ is a constant independent of $m$.
\end{lemma}
\begin{proof}
Fig. \ref{FIG:worst_sinr} illustrates the worst interference scenario of Phase $2$, where the groups in the shaded cells denote the active groups, determined by $4$-TDMA, and a single transmission pair is served for given time in each active group.
Let $d=n^{-q/2}$ be the length of a cell.
To obtain a lower bound on $C(k,i,j)$, we assume that there exist $8$ interferers at distance $d$ from the receiver, $16$ interferers at distance $3d$, $32$ interferers at distance $5d$, and so on.
Then $C(k,i,j)$ is lower bounded by
\begin{eqnarray}
C(k,i,j)\!\!\!\!\!\!\!\!&&\geq\frac{1}{n_2}\log\left(\frac{P_1(\sqrt{2}d)^{-\alpha}}{1+P_1\sum_{i=1}^{\infty} 8i((2i-1)d)^{-\alpha}}\right)\nonumber\\
&&\geq\frac{1}{n_2}\log\left(\frac{P_1}{(\sqrt{2}d)^{\alpha}+2^{\alpha/2+3}P_1\sum_{i=1}^{\infty}i^{1-\alpha}}\right),
\end{eqnarray}
where $\frac{1}{n_2}$ comes from $n_2$-TDMA between transmission pairs in the same scheduling set.
The first inequlity holds since we assume infinity number of interferers and the second inequality holds since we assume more interferers such that there are $8$ interferers at distance $d$, $16$ interferers at distance $2d$, and so on.
Notice that $(\sqrt{2}d)^{\alpha}\to0$ as $m\to\infty$ and $\sum_{i=1}^{\infty}i^{1-\alpha}=\zeta(\alpha-1)$, where $\zeta(s)\triangleq \sum_{i=1}^{\infty}i^{-s}$ denotes the Riemann zeta-function which converges to a finite value if $s>1$.
Thus there exists a positive constant $C_1$ satisfying $C(k,i,j)\geq C_1 n_2^{-1}$, which completes the proof.
\end{proof}

\begin{lemma} \label{THM:rate_p2_hier}
Suppose that the hierarchical cooperation is used within a group.
For sufficiently large $m$, $C(k,i,j)$ is lower bounded by $C_2n_2^{-\epsilon}$ whp, where $C_2>0$ is a constant independent of $m$ and $\epsilon>0$ is an arbitrarily small constant.
\end{lemma}
\begin{proof}
We refer Theorem  3.2 in \cite{OzgurLevequeTse:06} that all nodes can transmit with rate $C_2n_2^{-\epsilon}$ whp by the hierarchical cooperation.
\end{proof}

The following theorem shows the scaling behavior of $N\times M$ MIMO in which the channel matrix has $i.i.d.$ elements.
For our far-field channel, on the other hand, the channel elements are not $i.i.d.$ due to the different path-loss terms.
Thus we further lower the achievable rate in (\ref{EQ:R_kj}) to make the channel elements $i.i.d$. and apply the result of the following theorem.

\begin{theorem} [Lozano and Tulino] \label{THM:lozano}
Consider the ergodic capacity of $N\times M$ MIMO channel given by $C=\mathbb{E}\left(\log\det\left(\mathbf{I}_N+\frac{P}{M}\mathbf{H}\mathbf{H}^{\dagger}\right)\right)$, where the elements of $\mathbf{H}$ are $i.i.d$ with unit variance.
Define $a\triangleq\frac{M}{N}$.
As $M$ and $N$ tend to infinity, $\frac{C}{N}$ scales whp as
\begin{equation}
\begin{cases}\log(1+P) &\mbox{ if }a\to\infty\\
2\log \left(\frac{1+\sqrt{1+4P}}{2}\right)-\frac{\log e}{4P}\left(\sqrt{1+4P}-1\right)^2 &\mbox{ if }a\to1\\
a\log\left(\frac{P}{a}\right)+\operatorname{O}(a) &\mbox{ if }a\to 0\\
\end{cases}
\end{equation}
\end{theorem}

\begin{proof}
We refer the readers \cite{Lozano:02} for the proof.
\end{proof}

\subsubsection{TDMA-based cooperation}
Consider the sum-rate scaling when the TDMA-based cooperation is applied within each group.
\begin{theorem} \label{THM:rate_scaling1}
Suppose the single-source wireless network.
If the network performs the quantize-and-forward-based cooperative MIMO using the TDMA-based cooperation, for sufficiently large $m$
\begin{equation}
R_{\operatorname{sum}}=\begin{cases}\Theta\left(m\log m\right) &\mbox{ if }\beta>2\\
\Theta\left(m^{\beta/2}\right)  &\mbox{ if }0<\beta\leq2
\label{EQ:achiebavle_TDMA}
\end{cases}
\end{equation}
is achievable whp, where we set $\Delta=\frac{1}{2}$ and $q=\frac{1}{2}$.
\end{theorem}
\begin{proof}
From (\ref{EQ:Quan_noise}) with Lemmas \ref{THM:max_pwr} and \ref{THM:rate_p2_tdma}, we obtain whp that
\begin{equation}
N^k_{ij}\leq\frac{P_0(1+\epsilon_p)+1}{2^{\frac{1-\Delta}{\Delta}\frac{C_1}{4}\frac{n}{n^2_2}}-1}\leq \frac{P_0(1+\epsilon_p)+1}{2^{\frac{C_1}{4(1+\delta)^2}}-1}\triangleq N_{Q,1}
\end{equation}
where the second inequality holds since $n_2\leq(1+\delta)n^{1-q}$ whp and $\Delta=q=\frac{1}{2}$.
Thus, for sufficiently large $m$, $R(k,j)$ in (\ref{EQ:R_kj}) is lower bounded whp as
\begin{eqnarray}
R(k,j)\!\!\!\!\!\!\!\!\!&&\geq\frac{1}{2n}\mathbb{E}\left(\log\det\left(\mathbf{I}_{n_2}+\frac{P_0}{m(1+N_{Q,1})}\mathbf{\Gamma}^k_0\mathbf{\Theta}^k_0\mathbf{\Theta}^{k\dagger}_0\mathbf{\Gamma}^k_0\right)\right)\nonumber\\
&&\geq\frac{1}{2m^{\beta/2}}\frac{1}{m^{\beta/2}}\mathbb{E}\left(\log\det\left(\mathbf{I}_{n_2}+\frac{P_0}{m(1+N_{Q,1})}\mathbf{\Theta}^k_0\mathbf{\Theta}^{k\dagger}_0\right)\right)\nonumber\\
&&\triangleq\frac{1}{2m^{\beta/2}}R'(k,j),
\label{EQ:r_kj_low}
\end{eqnarray}
where the second inequality holds since $[\mathbf{\Gamma}^k_0]_{ii}=\frac{\|\mathbf{r}_0-\mathbf{r}^k_{n_2}\|^{\frac{\alpha}{2}}}{\|\mathbf{r}_0-\mathbf{r}^k_i\|^{\frac{\alpha}{2}}}\geq 1$ for all $i\in\{1,\cdots, n_2\}$.
Now consider the rate scaling of $R'(k,j)$.
From Theorem $\ref{THM:lozano}$, we know $a=\frac{m}{n_2}\in[(1+\delta)^{-1}m^{1-\beta/2}, (1-\delta)^{-1}m^{1-\beta/2}]$ and $P=\frac{P_0}{1+N_{Q,1}}$.
If $\beta>2$, that is $a\to 0$ as $m\to\infty$, $R'(k,j)$ scales as $\Theta(a\log (\frac{P}{a}))$ whp, which shows $\Theta(m^{1-\beta/2}\log m)$ scaling.
If $\beta=2$ or $a=1$, $R'(k,j)$ scales as $\Theta(1)$ whp.
Finally if $0<\beta<2$ or $a\to \infty$, $R'(k,j)$ scales as $\Theta(1)$ whp.
Thus $R(k,j)$ scales whp as $\Theta(m^{1-\beta}\log m)$ if $\beta>2$ and $\Theta(m^{-\beta/2})$ if $0<\beta\leq 2$.
Since this scaling result holds for all $k$ and $j$, we obtain whp
\begin{equation}
R_{\operatorname{ind}}=\begin{cases}\Theta\left(m^{1-\beta}\log m\right) &\mbox{ if }\beta>2\\
\Theta\left(m^{-\beta/2}\right)  &\mbox{ if }0<\beta\leq2.
\end{cases}
\end{equation}
From the fact that $R_{\operatorname{sum}}=m^{\beta}R_{\operatorname{ind}}$, we obtain (\ref{EQ:achiebavle_TDMA}), which completes the proof.
\end{proof}

For the ad hoc network, the hierarchical cooperation is essential to achieve the capacity scaling.
For the single-source network, on the other hand, if $\beta>2$ the simple TDMA-based cooperation is enough to achieve the capacity scaling.
It suggests that if we deploy a relatively large number of destinations in the network, that is $\beta>2$, we can reduce the network complexity drastically while still achieving the sum-rate proportional to the number of source antennas.
As shown later, if $0<\beta\leq 2$, the hierarchical cooperation is required to achieve the capacity scaling.

\subsubsection{Hierarchical cooperation}
Consider the sum-rate scaling when the hierarchical cooperation is applied within each group.
\begin{theorem} \label{THM:rate_scaling2}
Suppose the single-source wireless network.
If the network performs the quantize-and-forward-based cooperative MIMO using the hierarchical cooperation, for sufficiently large $m$,
\begin{equation}
R_{\operatorname{sum}}=\begin{cases}\Theta\left(m\log m\right) &\mbox{ if }\beta> (1-\epsilon)^{-1}\\
\Theta\left(m^{\beta(1-\epsilon)}\right) &\mbox{ if }0<\beta\leq (1-\epsilon)^{-1}
\label{EQ:achiebavle_hier}
\end{cases}
\end{equation}
is achievable whp, where we set $\Delta=\frac{1}{2}$ and $q=\epsilon$ and $\epsilon>0$ is an arbitrarily small constant.
\end{theorem}
\begin{proof}
Since the overall procedure is similar to the proof of Theorem \ref{THM:rate_scaling1}, we briefly explain the outline of the proof.
From (\ref{EQ:Quan_noise}) with Lemmas \ref{THM:max_pwr} and \ref{THM:rate_p2_hier}, we obtain whp that
\begin{equation}
N^k_{ij}\leq\frac{P_0(1+\epsilon_p)+1}{2^{\frac{1-\Delta}{\Delta}\frac{C_2}{4}\frac{n^{1-\epsilon}}{n_2}}}\leq \frac{P_0(1+\epsilon_p)+1}{2^{\frac{C_2}{4(1+\delta)}}-1}\triangleq N_{Q,2}.
\end{equation}
For sufficiently large $m$,
\begin{equation}
R(k,j)\geq\frac{1}{2m^{\beta\epsilon}}\frac{1}{m^{\beta(1-\epsilon)}}\mathbb{E}\left(\log\det\left(\mathbf{I}_{n_2}+\frac{P_0}{m(1+N_{Q,2})}\mathbf{\Theta}^k_0\mathbf{\Theta}^{k\dagger}_0\right)\right),
\end{equation}
whp.
Since $a\in\left[(1+\delta)^{-1}m^{1-\beta(1-\epsilon)},(1-\delta)^{-1}m^{1-\beta(1-\epsilon)}\right]$ and $P=\frac{P_0}{1+N_{Q,2}}$, from Theorem $\ref{THM:lozano}$, we obtain whp $R_{\operatorname{ind}}=\Theta(m^{1-\beta}\log m)$ for $\beta>(1-\epsilon)^{-1}$, which is the case that $a\to 0$.
Similarly, we obtain whp $R_{\operatorname{ind}}=\Theta(m^{-\beta\epsilon})$ for $\beta\leq(1-\epsilon)^{-1}$.
From the fact that $R_{\operatorname{sum}}=m^{\beta}R_{\operatorname{ind}}$, we obtain (\ref{EQ:achiebavle_hier}), which completes the proof.
\end{proof}

Notice that, for $\beta>(1-\epsilon)^{-1}$, the sum-rate capacity of the single-source network scales as $\Theta(m\log m)$ whp.
For $0<\beta\leq(1-\epsilon)^{-1}$, the lower and upper bounds show a gap of $\Theta(m^{\epsilon}\log m)$ but it is negligible compared to $\Theta(m^{\beta})$ in the limit of large $m$.


\subsection{Multiple Antennas at Destinations}
Consider the effect of multiple antennas at the destinations.
Let $l=m^{\gamma}$ be the number of antennas at each destination, where $\gamma\in[0,1]$.
Then the upper bound is given by $R_{\operatorname{sum}}=\operatorname{O}(m^{\min\{1,\beta+\gamma\}}\log m)$, which is directly obtained from Theorem \ref{THM:rate_upper} by regarding each receive antenna as a node.
If we apply the TDMA-based cooperation, the rate of Phase $2$ scales as $\Theta(\frac{l}{ln_2})=\Theta(\frac{1}{n_2})$ since the rate increases proportionally to $l$ and there exist $ln_2$ receive antennas in a group, meaning $ln_2$-TDMA is needed in a scheduling set.
Thus $R_{\operatorname{sum}}$ scales whp as $\Theta(m\log m)$ for $\beta>2(1-\gamma)$ and $\Theta(m^{\beta/2+\gamma})$ for $\beta\leq 2(1-\gamma)$.
If we apply the hierarchical cooperation, the rate of Phase $2$ is the same as Lemma \ref{THM:rate_p2_hier}.
Thus $R_{\operatorname{sum}}$ scales whp as $\Theta(m\log m)$ for $\beta>(1-\gamma)(1-\epsilon)^{-1}$ and $\Theta(m^{\beta(1-\epsilon)+\gamma})$ for $\beta\leq(1-\gamma)(1-\epsilon)^{-1}$.
We omit the detailed procedure, which is the same as Theorems \ref{THM:rate_scaling1} and \ref{THM:rate_scaling2}.

In conclusion, if multiple antennas are equipped at each destination as well as the source, then linear scaling is achievable if $\alpha+\beta\geq 1$ and TDMA-based cooperation can achieve the optimal scaling if $\beta>(1-\gamma)(1-\epsilon)^{-1}$.

\section{Conclusion} \label{sec:conclusion}
In this paper, we consider the capacity scaling of the single-source wireless network when the source has $m$ antennas.
We propose the cooperative MIMO scheme using quantize-and-forward.
We show that, like the single-destination network, the sum-rate proportional to $m$ is achievable if $n=\Omega(m)$ even without CSIT.
The sum-rate capacity scales whp as
\begin{equation}
\begin{cases} \Theta(m\log m)&\mbox{ if } \beta>(1-\epsilon)^{-1}\\
\Omega(m^{\beta(1-\epsilon)}) \mbox{ and } \operatorname{O}(m^{\min\{\beta,1\}}\log m)&\mbox{ if } 0<\beta\leq(1-\epsilon)^{-1}.
\end{cases}
\end{equation}
Note that the gap between upper and lower bounds becomes negligible for sufficiently large $m$.
To achieve linear capacity scaling in $m$, we apply the hierarchical cooperation to relay the quantized received signals.
As the number of destinations becomes large enough to satisfy $\beta>2$, the simple TDMA-based cooperation also achieves the capacity scaling.





\newpage

\begin{table}
\caption{Summary of notations: $\mathbf{A}$ and $\mathbf{a}$ denote a matrix and vector, respectively}
\label{Table:simbols}
\begin{equation*}
    \begin{array}{|c|c|}
    \hline
    \mathbf{A}^{\dagger} (\mbox{ or }\mathbf{a}^{\dagger})& \mbox{Conjugate transpose of }\mathbf{A}(\mbox{ or } \mathbf{a})\\
    \hline
    \mathbf{A}^{T} (\mbox{ or }\mathbf{a}^{T})& \mbox{Transpose of }\mathbf{A}(\mbox{ or } \mathbf{a})\\
    \hline
    \operatorname{Tr}(\mathbf{A})& \mbox{Trace of }\mathbf{A}\\
    \hline
    [\mathbf{A}]_{ij}& \mbox{$(i,j)$-th component of }\mathbf{A}\\
    \hline
    \mathbf{I}_n& \mbox{$n\times n$ identity matrix}\\
    \hline
    \|\mathbf{a}\|& \mbox{norm of $\mathbf{a}$}\\
    \hline
    \diag(a_1,\cdots,a_n)& \mbox{diagonal matrix with $[\diag(a_1,\cdots,a_n)]_{ii}=a_i$}\\
    \hline
    \end{array}
\end{equation*}
\end{table}


\end{document}